\documentclass[a4paper, 10pt, conference]{ieeeconf}      

\makeatletter
\let\NAT@parse\undefined
\makeatother
\usepackage[numbers]{natbib}
\IEEEoverridecommandlockouts                              
\title{\LARGE \bf
Portable Trust: \\
biometric-based authentication and blockchain storage for \\
self-sovereign identity systems
}

\usepackage{tabulary}
\usepackage{graphicx}
\usepackage[colorinlistoftodos]{todonotes}
\usepackage{acronym}
\usepackage{a4wide}
\usepackage{graphicx}
\usepackage{caption}
\usepackage{subcaption}
\usepackage{wrapfig}

\usepackage{fancyhdr}
\fancypagestyle{plain}{
  \fancyhf{}
  \fancyhead[C]{Conference on \LaTeX}     
  \fancyfoot[L]{This is a notice}

}

\author{
J.S. Hammudoglu, J. Sparreboom, J.I. Rauhamaa, J.K. Faber, L.C. Guerchi,\\
I.P. Samiotis, S.P. Rao and J.A. Pouwelse (course supervisor).\\
Computer Science department, Delft University of Technology, The Netherlands.\\
 \\
\textemdash~student project~\textemdash
}

\begin{document}

\maketitle
\thispagestyle{empty}
\pagestyle{empty}
\begin{abstract}

We devised a mobile biometric-based authentication system only relying on local processing. Our Android open source solution explores the capability of current smartphones to acquire, process and match fingerprints using \emph{only} its built-in hardware. Our architecture is specifically designed to run completely locally and autonomously, not requiring any cloud service, server, or permissioned access to fingerprint reader hardware. It involves three main stages, starting with the fingerprint acquisition using the smartphone camera, followed by a processing pipeline to obtain minutiae features and a final step for matching against other locally stored fingerprints, based on Oriented FAST and Rotated BRIEF (ORB) descriptors. We obtained a mean matching accuracy of 55\%, with the highest value of 67\% for thumb fingers. 
Our ability to capture and process a finger fingerprint in mere seconds using a smartphone makes this work usable in a wide range of scenarios, for instance, offline remote regions.
This work is specifically designed to be a key building block for a self-sovereign identity solution and integrate with our permissionless blockchain for identity and key attestation.
\end{abstract}

\begin{figure*}[!t]
    \centering
    \includegraphics[height=2.1in]{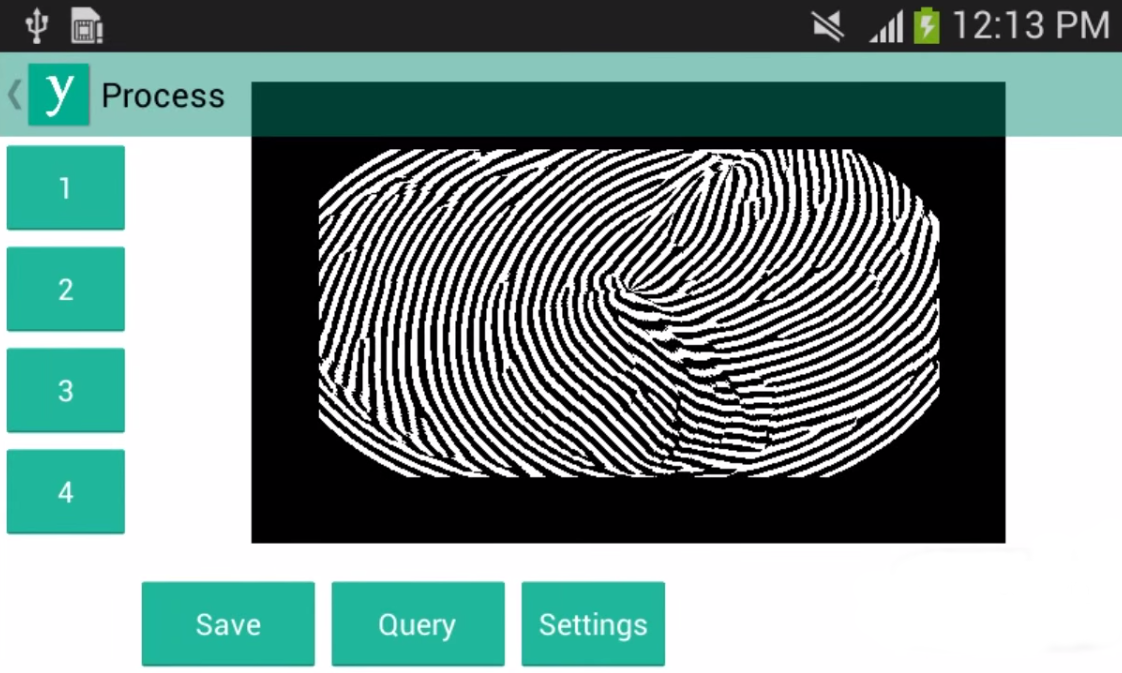}
    \caption{Main inspiration of this work, from University of Southampton (copied from~\cite{noureldienhussein})}
     \label{fig:fingerprintrecognition}
\end{figure*}

\section{INTRODUCTION} \label{intro}
Our aim is to create an initial step towards portable trust, in our narrow focus defined as authentication of trustworthy individuals with minimal constraints on mobility.

We created a fully functional end-to-end architecture for biometric-based authentication solution for offline usage and devised an operational prototype.
Permissionless and autonomous systems which do not rely on central servers or even Internet access are challenging to create.
Using only local processing is especially difficult.

Fingerprint authentication is a reliable biometric method used for identification of individuals in governments, organizations and  public \& private institutions for several purposes such as custom and border protection, suspect identification, access control to restricted areas, assets and information. Nowadays, most of the systems developed for this purpose work either as standalone applications using specialized hardware, or in a centralized architecture. The aim of this research is to break that barrier and demonstrate the viability of a \emph{portable trust system}, cutting out the middlemen, taking advantage of Android smartphones to acquire, process and match fingerprint minutiae.

We believe our operational prototype is one of the rare examples of a complete open source end-to-end solution. The main reasons for using fingerprints as our biometrics are the vast pre-existing research that has been conducted on the subject and its popularity. We also wanted to explore the possibility to use a smartphone camera for fingerprint acquisition. 
Our motivation for using the build-in camera is that it can be freely used, no permission or contractual agreement with the manufacturer is required. This in contrast to fingerprint scanners, which are not included on all smartphones, require special permissioned access, and can not be freely used by the software.

Self-sovereign identity solutions empower users to locally store all their personal information and trustworthy public keys. This mean users do not need to trust identity providers, they take responsibly, and store data securely without any risk of privacy abuse~\cite{self-sov-ID}.
This permissionless and autonomous approach to identity is a natural match for our biometric-based authentication architecture. Experimental state-of-the-art self-sovereign identity system already employ blockchain technology~\cite{baars2016towards}. 
Our implementation aims to be permissionless, open source and guarantee interoperability across its layers, facilitating our ongoing integration with blockchain-based identity systems~\cite{tribler-wiki}.
\section{RELATED WORK} 

\begin{figure}[!h]
    \centering
    \includegraphics[height=3.05in]{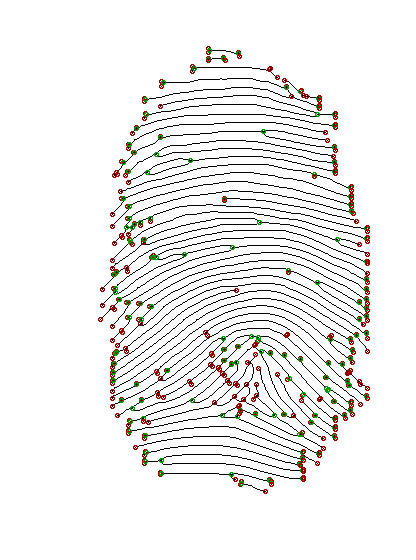}
    \caption{Minutiae extraction from skeletonized image (copied from~\cite{wrocklaw})}
     \label{fig:minutiae}
\end{figure}

Several approaches have been developed for fingerprint authentication using Android smartphones. Even though these resources have often been limited to legal patents as in the case of Scale Invariant Feature Transform (SIFT)  algorithm~\cite{lowe1999object} and specific or additional hardware like built-in fingerprint scanners, they have served as a guide for development of the proposed solution. \citet{RxFingerprint} has examined issues related to secure user information via fingerprint authentication using Android Fingerprints API \cite{AndroidAPI}. This approach is not suitable for our project because the API does not allow sharing of fingerprints, however the encryption and decryption system using AES and RSA standards could be used as a guide to increase security. Our image processing pipeline is based on the one described in the Fingerprint Recognition project~\cite{noureldienhussein}, shown in figure~\ref{fig:fingerprintrecognition}. It is an Android application that uses smartphone camera to capture an image for preprocessing, skeletonization and feature extraction. The way our algorithm extracts minutiae from the skeletonized image is based on an algorithm presented by University of Wrocklaw~\cite{wrocklaw} 
where black pixels are considered as bifurcation points if they have three black neighbours and as ridge endings if they have one black neighbour, as shown in figure~\ref{fig:minutiae}.

\section{ALGORITHM \& IMPLEMENTATION OF FINGERBLOX}

The algorithm is divided into three major components responsible for the fingerprint acquisition, the image processing pipeline and fingerprint matching. 

\subsection{Image acquisition}

Initially, the user opens the application and is prompted to take a picture of a fingerprint inside the elliptic area as shown in figure \ref{fig:home}. The elliptic area is used as a mask for the cropping step later, in the processing stage. 

\subsection{Image processing pipeline}
\label{ipp}

An ordered sequence of processing stages are applied to the captured image, starting with skin detection using HSV color space approach \cite{oliveira2009skin}, fingerprint isolation for the rest of the image applying a series of erosion and dilatation morphological operations, noise reduction using a Gaussian blur linear filter, and finally the fingerprint area is elliptically cropped from the rest of the image using a skin detection mask.

After this point further stages focus on enhancing and extraction of fingerprint minutiae features. As initial step, ridges (darker curves) and valleys (brighter curves) differentiation using contrast enhancements \cite{gao2001fingerprint}, a lightweight representation of the fingerprint is produced using Thinning(Skeletonization) applying the Zhang-Suen algorithm \cite{zhang1984fast} to  reduce ridges to one pixel. Lastly, ridge endings and bifurcation points are detected by iterating over pixels and classifying image locations in rectangular coordinates(x,y) as ridge endings when it has one white neighbor pixel and as bifurcations when it has three white neighbors. From detected minutiae ORB descriptors are calculated and stored to be used for matching purposes. ORB was selected because it is twice as fast as SIFT \cite{rublee2011orb} and not patented. Additionally, it is more suitable for low-power consumption devices such as smartphones. 

\subsection{Matching}

Fingerprint matching is done using minutiae keypoints and ORB descriptors obtained as described in the previous section and compared against the stored ones using random sample consensus (RANSAC) and ratio test \cite{lowe2004distinctive} to obtain the best matches. 

\subsection{Implementation}

Our implementation~\cite{fingerblox} is public, open source, and hosted on Github. It is written in Java as it's the native language for Android applications and uses the popular library Open Source Computer Vision (OpenCV)~\cite{opencv}. The line counts of source files are listed in table~\ref{lines-of-code}. Most of the code resides in ImageProcessing.java which contains the implementation for the image processing pipeline.

\begin{table}[]
\centering
\caption{List of source files by lines of code}
\label{lines-of-code}
\begin{tabular}{ll}
\textbf{Source file}      & \textbf{Lines of code} \\
\cline{1-2} 
ImageProcessing.java      & 1283                   \\
MainActivity.java         & 351                    \\
ImageDisplayActivity.java & 333                    \\
CameraView.java           & 110                    \\
ScannerOverlayView.java   & 82                     \\
CameraOverlayView.java    & 55                     \\
ImageSingleton.java       & 9                      \\
total                     & 2223                  \\
\cline{1-2} 
\end{tabular}
\end{table}



\section{User Experience}
\label{userexp}

\begin{figure}[!h]
    \centering
    \includegraphics[height=4.0in]{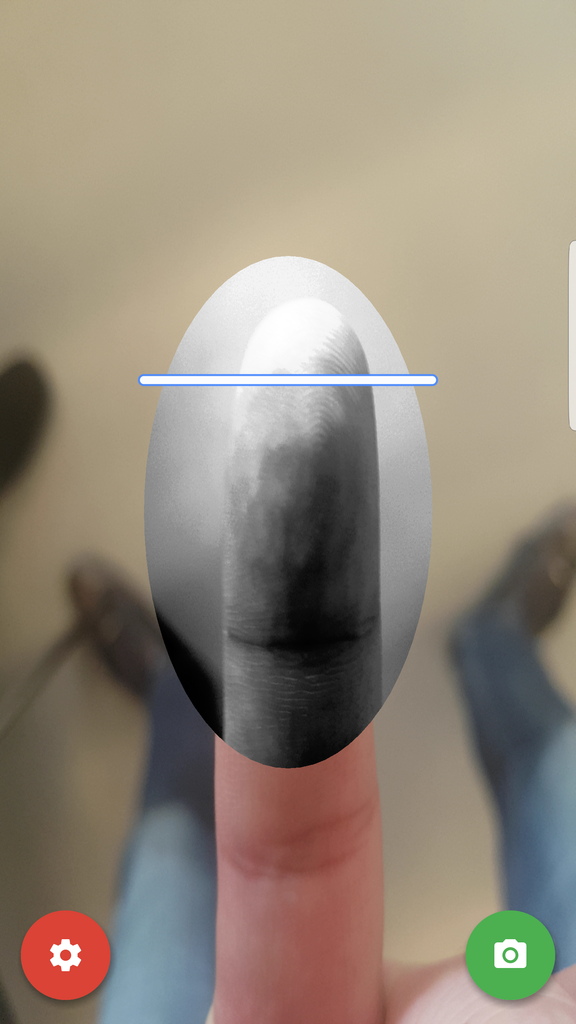}
    \caption{Main fingerprint acquisition}
     \label{fig:home}
\end{figure}

 
From the user experience perspective the application was divided into 3 screens to allow intuitive navigation. When it is started, an elliptical area in the middle of the screen shows the user where to place the finger as shown in figure \ref{fig:home}. The preview option allows the user to inspect partly processed image before taking the picture.

\begin{figure}[!h]
    \centering
    \includegraphics[height=4.0in]{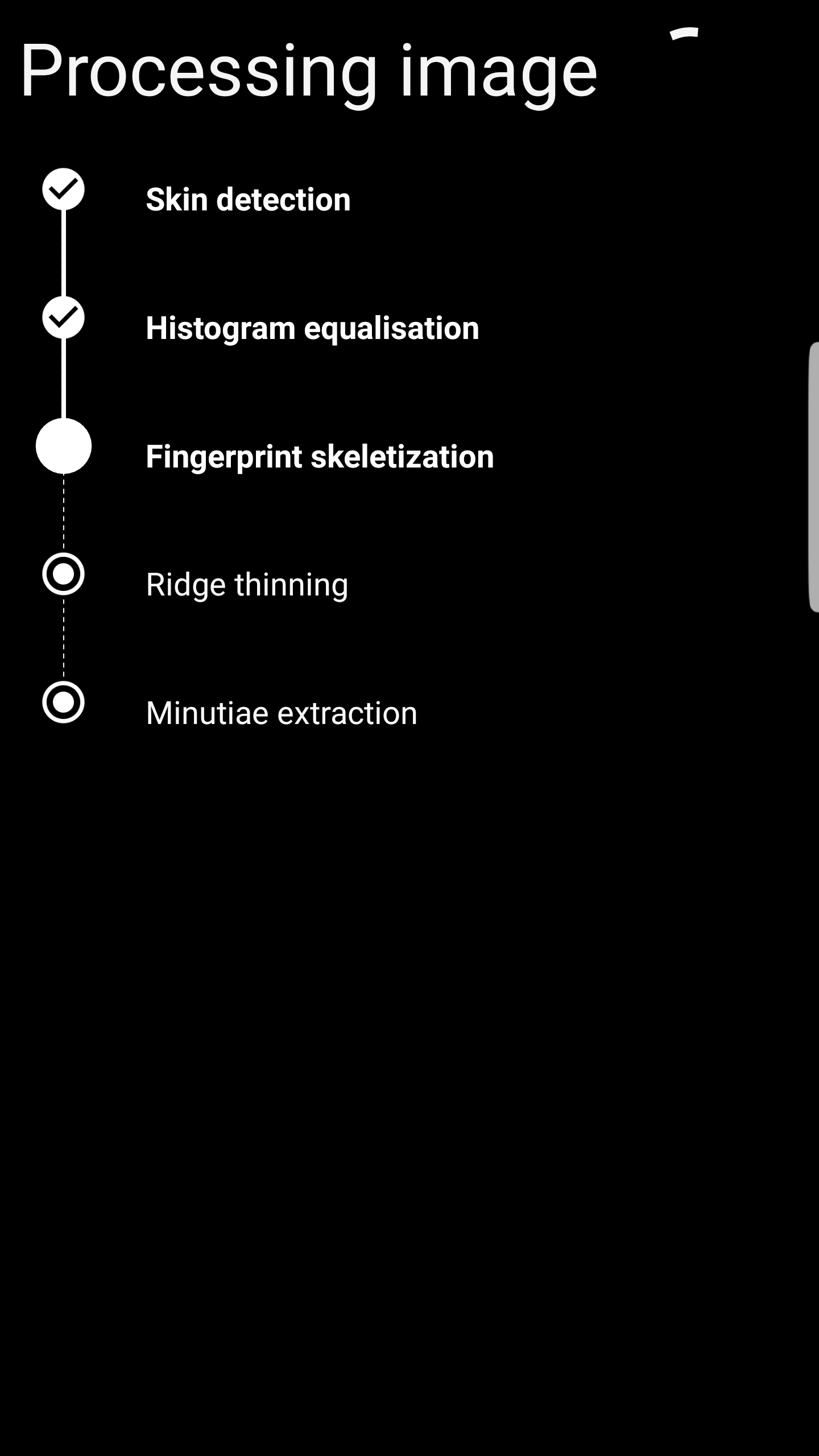}
	\caption{Processing pipeline display}
         \label{fig:pipeline}
\end{figure}

\begin{figure}[!h]
    \centering
    \includegraphics[height=4.0in]{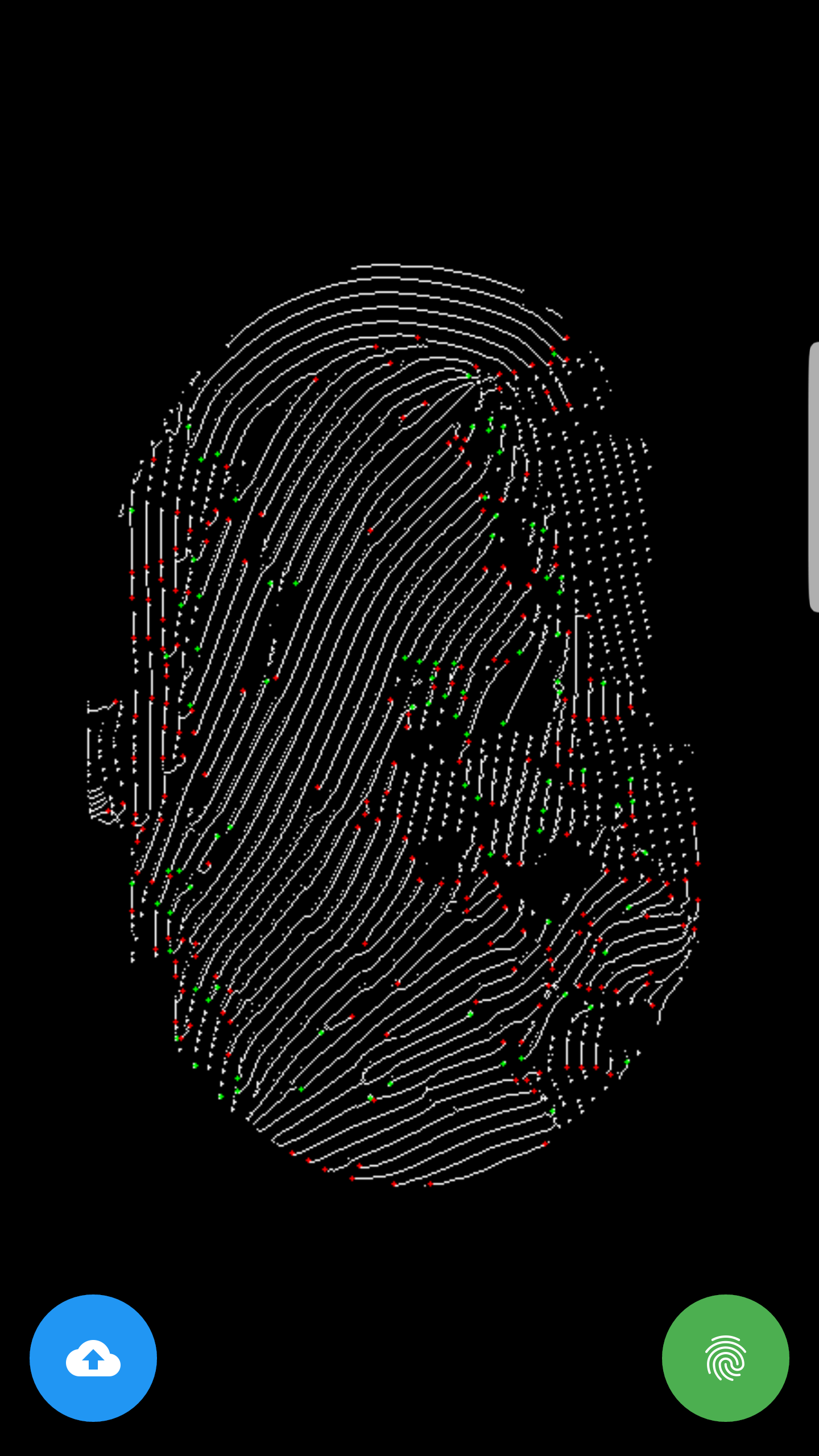}
    \caption{Feature extraction}
    \label{fig:minutiae}
\end{figure}


When the processing task starts, immediate updates of the current step are displayed as shown in figure \ref{fig:pipeline} to inform the user about the current status of processing the fingerprint. 

Finally, as shown in figure \ref{fig:minutiae}, the fingerprint features can be saved for future comparisons or compared against a local database of previously obtained minutiae data of fingerprints located in the phone's file system. The entire capture and matching process took less than one minute on a Samsung Galaxy S7 Edge smartphone.


\section{EVALUATION} \label{algorithm}

We conducted a small performance evaluation experiment. We measured the authentication accuracy using different lighting conditions and different background scene when compared to the reference image.

The performance of our system is evaluated against a small dataset of 20 finger images belonging to one of the authors. Of these 20 finger images, 10 belong to the same right hand, where 5 are used for extracting feature points and the other 5 for matching. It is similarly followed for the left hand too. The fingers have been numbered from one to five starting on the thumb finger and the prefixed L for left and R for right hand.

The matching algorithm gives the matching percentage between two images. This metric is currently used as a performance measure for the algorithm. A higher percentage indicates a higher confidence over that fingerprint matching.

\begin{table}[h]
\centering
\begin{tabular}{rrrrrr}
                     & \textit{\textbf{L1}} & \textit{\textbf{L2}} & \textit{\textbf{L3}} & \textit{\textbf{L4}} & \textit{\textbf{L5}} \\ \cline{2-6} 
\textit{\textbf{L1}} & \textbf{52}          & 3                    & 42                   & 11                   & 5                    \\  
\textit{\textbf{L2}} & 11                   & \textbf{56}          & 26                   & 14                   & 10                   \\
\textit{\textbf{L3}} & 7                    & 37                   & \textbf{59}          & 42                   & 4                    \\
\textit{\textbf{L4}} & 14                   & 17                   & 23                   & \textbf{67}                   & 9                    \\
\textit{\textbf{L5}} & 5                    & 4                    & 8                    & 4                    & \textbf{35}                   \\ \cline{2-6} 
\end{tabular}
\caption{Fingerprint matching accuracy for each finger of the left hand.}
\label{left}
\end{table}

\begin{table}[h]
\centering      
\begin{tabular}{rrrrrr}
                     & \textit{\textbf{R1}} & \textit{\textbf{R2}} & \textit{\textbf{R3}} & \textit{\textbf{R4}} & \textit{\textbf{R5}} \\ \cline{2-6} 
\textit{\textbf{R1}} & \textbf{67}                  & 5                    & 10                   & 10                   & 4                    \\  
\textit{\textbf{R2}} & 12                   & \textbf{54}                   & 34                   & 11                   & 6                    \\
\textit{\textbf{R3}} & 9                    & 26                   & \textbf{55}                   & 40                   & 9                    \\
\textit{\textbf{R4}} & 6                    & 10                   & 31                   & \textbf{58}                   & 5                    \\
\textit{\textbf{R5}} & 5                    & 8                    & 8                    & 9                    & \textbf{46}                   \\ \cline{2-6} 
\end{tabular}
\caption{Comparison of matching accuracies across each finger of the right hand.}
\label{right}
\end{table}

Table~\ref{left} and table~\ref{right} show the matching percentages of the fingers of the left and right hand respectively. The finger labels on the Y-axis represent the images that have been used for extracting feature points, while image labels on X-axis are the images against which the matching was performed.

The mean percentage of performance of our system was 55\%. It was calculated from the mean of the matching percentages represented in the diagonal values of table~\ref{left} and table~\ref{right}. 

Since our image processing pipeline was built to identify pictures of fingers taken with low-end devices as opposed to images of processed fingerprint skeletons, we could not perform evaluation on an available dataset of fingerprints like the NIST special database 27 \cite{garris2000nist}. The NIST dataset has gray-scale or binary images which would not be ideal when implementing our skin detection algorithm. 
Our small scale experiment is too limited to calculate other performance metrics like precision, recall and F1-scores with statistical accuracy. 
We believe our work shows the viability of completely offline, autonomous, and permissionless biometric-based authentication.

The size of the local storage limits the scalability of our solution.
The approximate storage requirement for a single matched fingerprint minutiae feature is 320 bits.
Processing of a typical fingerprint yields roughly 300~features per fingerprint. Therefore, devices such as the Google Pixel phone (128~GByte) can contain in the order of 10~million fingerprints.
Novel approaches are required to quickly search through inherently linear data-structures such as the blockchain~\cite{jin2012fingerprint}. 
Other recent work~\cite{} is focused on the privacy challenge, as fingerprint should be stored in encrypted form on any blockchain. 

\section{CONCLUSIONS}


Trust is central to society, but proven difficult to establish or grow.
Our work aims to create a first step towards portable trust, authentication of trustworthy individuals with minimal constraints on mobility.

We created a generic building block for authentication and identity management in general.
Our solution does not depend on special fingerprint hardware, cloud providers, identity issuers, central servers, middleman or even Internet connectivity.
We successfully created a biometric-based authentication prototype which is permissionless, autonomous, and open source.

We conducted a small 20-finger performance evaluation experiment. The image processing pipeline using OpenCV and advanced ORB extraction features produced a mean matching accuracy of 55\% with a maximum 67\% for left hand thumb and minimum of 35\% for the left pinky finger.
This wide margin in accuracy was mainly produced in the acquisition step due to the manual zooming and ambient luminosity.

Further research is required to improve image quality, taking into consideration the aforementioned parameters. Using automatic and scene-adaptive methods during the acquisition step to verify image quality is a key aspect for future work.
One idea is to apply new tone mapping operators \cite{kopf2007capturing} after the image capture to mitigate shadow effects and the inclusion of a Gabor filter following the guidelines of \citet{gottschlich2012curved} to increase matching accuracy.
Other aspects to take into consideration for future research are the addition of other biometrics to our system such as face-recognition to increment reliability while taking advantage of the Android camera, the possibility to use neural networks as it has proven to have potential \cite{jea2005minutia} and the integration with blockchain technologies due to its ability for tamper-proof encrypted fingerprint storage.

\bibliography{References}
\addtolength{\textheight}{-12cm}

\end{document}